\documentclass[aps,prb,twocolumn,floatfix,amssymb,superscriptaddress]{revtex4}

\usepackage{graphicx}
\usepackage{dcolumn}
\usepackage{bm}
\usepackage{amsmath}
\begin{document}

\newcommand{\bk}{{\bf k}}
\newcommand{\bp}{{\bf p}}
\newcommand{\bv}{{\bf v}}
\newcommand{\bq}{{\bf q}}
\newcommand{\tbq}{\tilde{\bf q}}
\newcommand{\tq}{\tilde{q}}
\newcommand{\bQ}{{\bf Q}}
\newcommand{\br}{{\bf r}}
\newcommand{\bR}{{\bf R}}
\newcommand{\bB}{{\bf B}}
\newcommand{\bA}{{\bf A}}
\newcommand{\bE}{{\bf E}}
\newcommand{\bj}{{\bf j}}
\newcommand{\bK}{{\bf K}}
\newcommand{\cS}{{\cal S}}
\newcommand{\vd}{{v_\Delta}}
\newcommand{\tr}{{\rm Tr}}
\newcommand{\kslash}{\not\!k}
\newcommand{\qslash}{\not\!q}
\newcommand{\pslash}{\not\!p}
\newcommand{\rslash}{\not\!r}
\newcommand{\bs}{{\bar\sigma}}

\title{Quantum oscillations from Fermi arcs}

\author{T. Pereg-Barnea}
\affiliation{Department of Physics,
California Institute of Technology, 1200 E. California Blvd, MC114-36,
Pasadena, CA 91125 }
\author{H. Weber}
\affiliation{Department of Physics and Astronomy, University of British Columbia,
Vancouver, BC, Canada V6T 1Z1}
\affiliation{Institut f\"ur Theoretische Physik, Universit\"at zu K\"oln, Z\"ulpicher Str. 77, 50937 K\"oln, Germany}
\author{G. Refael}
\affiliation{Department of Physics,
California Institute of Technology, 1200 E. California Blvd, MC114-36,
Pasadena, CA 91125 }
\author{M. Franz}
\affiliation{Department of Physics and Astronomy,
University of British Columbia, Vancouver, BC, Canada V6T 1Z1}

\date{\today}

\begin{abstract}
When a metal is subjected to strong  magnetic field $B$ nearly all
measurable quantities exhibit oscillations periodic in $1/B$. Such
quantum oscillations represent a canonical probe of the defining aspect of
a metal, its Fermi surface (FS). In this study we establish a new
mechanism for quantum oscillations which requires only finite segments
of a FS to exist. Oscillations periodic in $1/B$ occur if the FS segments
are terminated by a pairing gap.
Our results reconcile the recent breakthrough experiments showing
quantum oscillations in a cuprate superconductor YBa$_2$Cu$_3$O$_{6.51}$,
with a well-established result of many angle resolved photoemission
(ARPES) studies which consistently indicate ``Fermi arcs'' --
truncated segments of a Fermi surface -- in the normal state of the cuprates.
\end{abstract}
\maketitle

In conventional metals superconductivity can be understood as a
pairing instability of the Fermi surface \cite{bcs}. Recent
unambiguous identification of Shubnikov - de Haas \cite{taillefer1,bangura1}
and de Haas - van Alphen\cite{jaudet1} oscillations in
YBa$_2$Cu$_3$O$_{6.51}$ (YBCO) in high magnetic fields ushered a new
era in the field by furnishing a long awaited proof that a Fermi
surface exists in a high-temperature superconductor (SC). This
important discovery, which according to the conventional paradigm
implies a closed Fermi surface (FS), creates interesting new puzzles.
The existence of such a closed FS contradicts the well-established
result of many angle resolved photoemission (ARPES) studies which
consistently indicate ``Fermi arcs'' -- truncated segments of a Fermi
surface -- in the normal state of cuprates\cite{ding1,kanigel1,shen1}.
In this theoretical study we establish a new mechanism for quantum
oscillations which requires only finite segments of the FS to exist.
We present arguments that this new mechanism is relevant to the
cuprates and show that it accounts for the quantum oscillations in a
model that exhibits genuine Fermi arcs terminated by a pairing gap,
consistent with the ARPES data.

Quantum oscillation experiments in YBCO \cite{taillefer1,bangura1,jaudet1},
when analyzed using the conventional Onsager-Lifshitz picture
\cite{onsager1,lifshitz1,shoenberg1}, indicate small Fermi pockets each with
an area covering approximately 1.9\% of the first Brillouin zone (BZ). Such small
Fermi pockets do not arise naturally from the
band structure calculations and their total area is inconsistent
with the nominal doping concentration. To explain these results
proposals for various states with broken translational symmetry have been
put forward \cite{millis1,kee1,rice1,sachdev-galitski,podolski}, leading to complicated band structures
with multiple Fermi pockets. One would, however, expect to see signatures of
such a Fermi surface reconstruction in other experiments,
most notably the ARPES, which is capable of mapping out Fermi surfaces of
metals with high accuracy. Yet, extensive ARPES studies on various cuprate
materials show no evidence for Fermi pockets.

One can attempt to reconcile the quantum oscillation data with ARPES
by postulating that Fermi pockets do exist but cannot be seen by ARPES
for various reasons. Of these, the suppression of the photoemission
intensity on the pocket's back side due to the coherence factors is
often cited\cite{stanescu}.

In this study we adopt a radically different point of view -- we
assume that the Fermi arcs observed in ARPES are {\em real} and ask if
such genuine Fermi arcs can give rise to quantum oscillations.  With
one additional experimentally motivated assumption \cite{norman3},
namely that the gap terminating the arcs is of the pairing origin, we
find the answer to be affirmative. Our reasoning that underlies this
conclusion is a natural extension of the conventional Onsager-Lifshitz
picture supplemented by the analysis of what happens once the electron
reaches the arc endpoint.  These semiclassical considerations are then
supplemented and confirmed by exact numerical diagonalizations of a
fully quantum lattice model.

In what follows we advance two principal ideas. First, we formulate a
simple, experimentally motivated model for a normal state of
underdoped cuprate superconductors that exhibits genuine Fermi arcs
terminated by a pairing gap. Second, we demonstrate that in an applied
external magnetic field the low-energy density of states (DOS) in this
model oscillates as a function of energy as well as magnetic field.
The DOS is periodic in energy with frequency that is linear in $1/B$ and
proportional to the Fermi arc length.  At the Fermi level (or any
other fixed low energy) the DOS oscillates with $1/B$.  The frequency of
oscillations is not related to any area in momentum space; instead it
is proportional to the gap amplitude and the fraction of the FS that
is gapped.
The origin of the quantum oscillations in our model is the periodic
appearance of low energy Andreev-type bound states\cite{adagideli} associated with the fermi arcs and is distinct from all mechanisms proposed to explain quantum oscillations in cuprates in the existing literature\cite{millis1,kee1,rice1,sachdev-galitski,podolski,varma,vafek}.
\begin{figure*}
\includegraphics[width = 8.0cm]{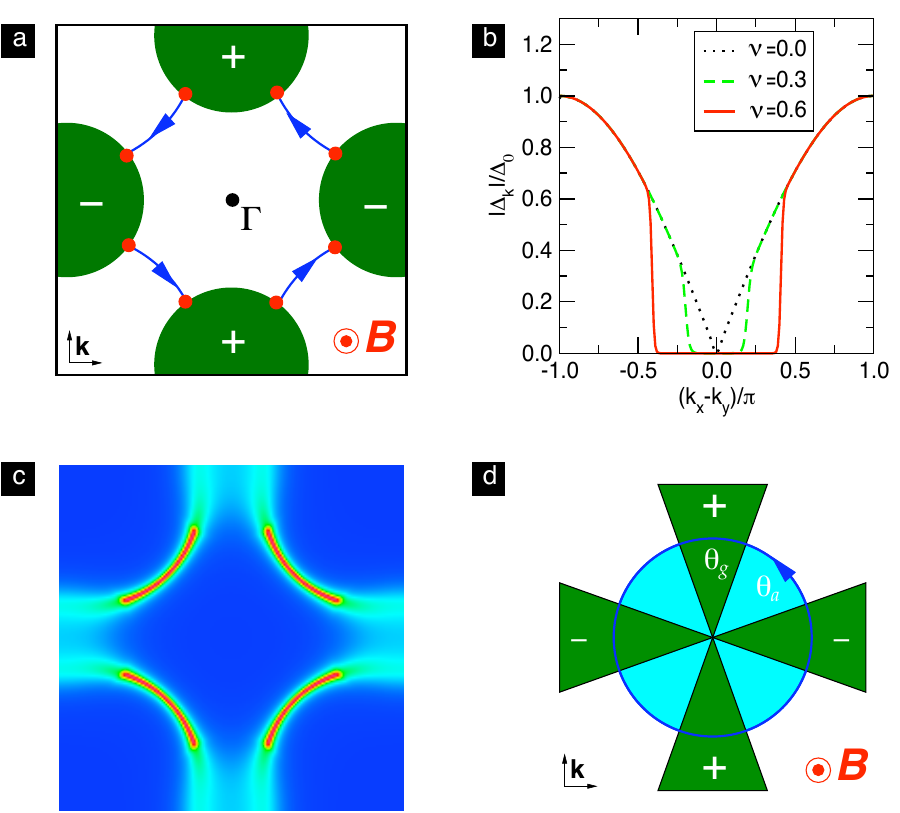}
\caption{{\bf The Fermi-arc metal. }
a) The gap structure of the Fermi arc metal (FAM) in the shaded areas
adjacent to the BZ faces the gap is nonzero while the excitations in
the rest of the BZ are gapless. Blue lines represent the gapless
segments of the Fermi surface known as the `Fermi arcs'. When a
perpendicular magnetic field $B$ is applied to the system electrons in
the gapless regions follow semiclassical orbits which coincide with
contours of constant energy $\epsilon_\bk$. b) The amplitude of the modified $d$-wave
gap $\tilde\Delta_\bk$, Eq.\ (\ref{dk}), with $\tau=0.01$ plotted along
the line connecting points $(0,\pi)$ and $(\pi,0)$ of the BZ. c)
Spectral intensity, $A(\bk,\omega)=-\pi^{-1}{\rm Im}G(\bk,\omega)$, of
the Fermi-arc metal at $\omega=0$.  Here
$G(\bk,\omega)=[\omega-\epsilon_\bk+i\Gamma-\tilde\Delta_\bk^2/(\omega+\epsilon_\bk+i\Gamma)]^{-1}$
is the Green's function of Hamiltonian (\ref{h1}), with additional
impurity broadening represented by $\Gamma=25$meV.  We use the
tight-binding parametrization for $\epsilon_\bk$ of Ref.\
\onlinecite{norman0} and $\Delta_0=50$meV, $\nu=0.6$ and $\tau=0.01$.  d)
Simplified model of FAM with circular Fermi surface and piecewise
constant gap used in semiclassical analysis. We denote the
angular extent of the arcs and gapped regions as $\theta_a$ and
$\theta_g =\pi/2-\theta_a$ respectively.
}
\label{fig1}
\end{figure*}
%

\section{Fermi-arc metal}

We now provide a justification for the above claims. Our starting
point is a simple phenomenological model for the `Fermi-arc metal'
(FAM) that we take to describe the non-superconducting state of underdoped
 cuprates with the
quasiparticle excitation spectrum adiabatically connected to the
ordinary $d$-wave superconductor. The latter is well-known to describe
the superconducting ground state of cuprates. It has a pairing gap
$\Delta_\bk=\Delta_0\chi_\bk$ with $\chi_\bk={1\over 2}(\cos k_x-\cos
k_y)$ and quasiparticle excitation spectrum
\begin{equation}\label{ek}
E_\bk=\sqrt{\epsilon_\bk^2 + \Delta_\bk^2},
\end{equation}
with $\epsilon_\bk$ the band dispersion referenced to the Fermi level
$\epsilon_F$. We assume that the excitations of the Fermi-arc metal have the
same BCS form as Eq.\ (\ref{ek}) but with $\Delta_\bk$ replaced by a {\em
modified} $d$-wave gap $\tilde\Delta_\bk$ which {\em vanishes} along the BZ
diagonals as illustrated in Fig. \ref{fig1}a. Although the detailed form of
$\tilde\Delta_\bk$ is unimportant for the subsequent discussion we often find it
useful to parameterize it as
\begin{equation}\label{dk}
\tilde\Delta_\bk=\Delta_0{\chi_\bk\over e^{(\nu^2-\chi_\bk^2)/\tau^2}+1}.
\end{equation}
Here $\nu$ sets the length of the arc while $\tau$ controls the width of the
step between the large-gap and zero-gap regions as illustrated in Fig.\
\ref{fig1}b. The ordinary $d$-wave
superconductor is recovered in the limit $\nu,\tau\to 0$.

Several remarks are in order before we address quantum oscillations in
this model.
(i) We think of the above postulated Fermi-arc metal spectrum as originating
from a BCS-type pairing Hamiltonian
\begin{equation}\label{h1}
{\cal H}=\sum_{\bk\sigma}\epsilon_\bk c^\dagger_{\bk\sigma}c^{}_{\bk\sigma}
+\sum_{\bk}(\tilde\Delta^{}_\bk\ c^\dagger_{\bk\uparrow}c^{\dagger}_{-\bk\downarrow}+{\rm h.c.})
\end{equation}
where $c^\dagger_{\bk\sigma}$ creates an electron with momentum $\bk$ and
spin $\sigma$. Although the microscopic Hamiltonian that would stabilize
the mean-field state described by Eqs.\ (\ref{dk},\ref{h1}) is presently
not known, we see no fundamental reason why
such a state could not occur for suitably chosen electron-electron
interaction. We remark that Fermi arcs have been argued to appear in a
phase-fluctuating SC \cite{fm1,altman1} and various more exotic
quantum states such as the ``algebraic charge liquid''
\cite{kaul1}. (ii) A system described by Hamiltonian (\ref{h1}) would
in fact be a superconductor, albeit with low superfluid density due to
ungapped portions of the Fermi surface. We thus view this BCS
Hamiltonian as an effective mean-field theory valid on intermediate
length scales; at long length scales (compared to the magnetic length)
phase fluctuations disrupt long-range superconducting order and render
the system normal.  (iii) The motivation for our phenomenological
model comes primarily from experimental considerations. Indeed many
experimental studies hint at the existence of the residual
superconducting order in the pseudogap state
\cite{uemura1,corson1,pasler1,ong1,ong2,lem1}. A Bogoliubov-type
dispersion of quasiparticles in cuprates above $T_c$, observed very
recently by ARPES \cite{norman3}, provides further direct evidence for
the pairing nature of the pseudogap. The spectral intensity computed
from our model and displayed in Fig.\ \ref{fig1}c is in detailed
agreement with this data. Finally, (iv) we note that within the space
of mean-field models the above construction appears to be the only way
to construct genuine Fermi arcs. As mentioned above, particle-hole
instabilities without exception produce closed Fermi surfaces which
can only terminate at BZ boundaries.

\section{Semiclassical analysis}

\begin{figure*}
\includegraphics[width=17cm]{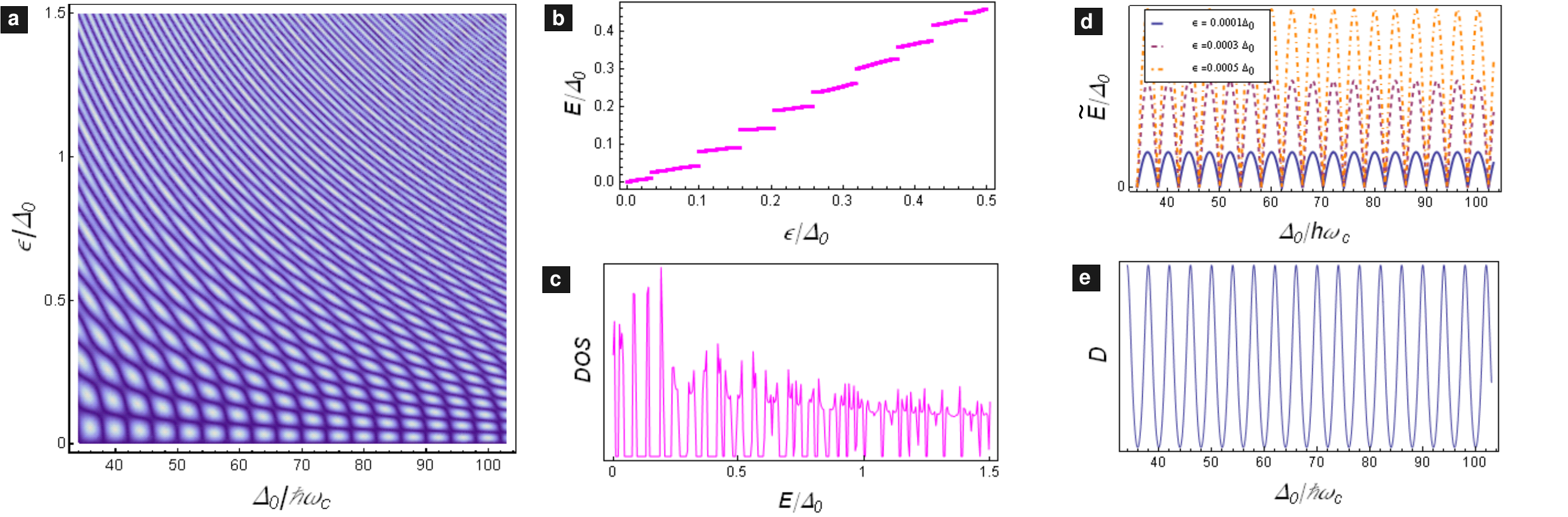}
\caption{{\bf Semiclassical analysis.}
a) Density plot of the quasi-energy $\tilde{E}$ as a
function of the
inverse magnetic field (times $\Delta_0 m_c c/e$) and the band
energy (in units of the gap amplitude). $\tilde{E}$ is
analytically obtained for the simplified Fermi-arc metal with a
piecewise constant gap structure; it is scaled by
$\hbar\omega_c$ and ranges between 0 (dark) and 1 (bright).  b)
Quasiparticle excitation energy, $E$, in the Fermi-arc metal as a function
of the band energy $\epsilon$ at constant $B$.
The quasi-particle energy is related to the quasi-energy
$\tilde{E}$ as $E=\tilde{E}+n\hbar\omega_c$, where
$n$, the unwinding index, is chosen to continuously match $E$ with
$\epsilon$ for $\epsilon\gg\Delta_0$, and maintain $E$ as a piecewise
continuous and monotonically increasing function of $\epsilon$.  c)
Density of states at a constant magnetic field $B\approx 40$T as a
function of energy.  d) The quasi-energy, $\tilde{E}$, for several low lying states
with $\epsilon<\hbar\omega_c$ vs. $1/B$.  e) The density of states at
the Fermi energy vs. the magnetic field.  In both plots the
given field range corresponds to $20-60$T.  The number of DOS oscillations
found in this range is about 17, in good agreement with the experiments.  In all panels $\theta_a = \pi/4$
corresponding to $\nu\sim 0.6$ in the numerical study.}\label{quasienergy}
\end{figure*}

In the absence of a gap, the classical equations of motion for an electron wave packet are
\begin{equation}\label{semi1}
\hbar\dot{\bk}=-{e\over c}(\bv_\bk\times\bB), \ \ \ \
\dot{\br}=\bv_\bk  = {1\over \hbar}\nabla_{\bk}\epsilon_{\bk}.
\end{equation}
Eqs.\ (\ref{semi1}) imply that the electrons move on constant
energy contours in momentum space.  The time it takes to complete a
cycle is  $T=2\pi/\omega_c$ with $\omega_c=eB/m_cc$ the
cyclotron frequency.

When constructing a wave packet to describe the semiclassical motion
in the FAM, clearly the motion on the arcs is governed by Eq.\
(\ref{semi1}).  However, the evolution in momentum space drives the
wave packet into the gapped region where its charge is no longer a
good quantum number due to electron-hole mixing induced by the pairing
gap.  We therefore choose to construct our semiclassical wave packet
out of `bogolons' rather than electrons. These are Bogoliubov-de
Gennes (BdG) quasiparticles of the underlying superconductor and can
be thought of as coherent mixtures of electrons and holes.  The bogolon
wavefunction $\Psi_\bk=(u_{\bk},v_{\bk})$ is an eigenstate of the BdG
Hamiltonian
\begin{equation}\label{bdg1}
{\cal H}_\bk = \begin{pmatrix}\epsilon_\bk & \tilde\Delta_\bk \\
\tilde\Delta_\bk & -\epsilon_\bk\end{pmatrix} = \epsilon_\bk\sigma_3 + \Delta_\bk\sigma_1,
\end{equation}
where $\sigma_j$ are the Pauli matrices acting in the particle-hole space.

The momentum ${\bk}$ of the bogolon wavepacket
continues to evolve according to Eq.\ (\ref{semi1}) since both the
velocity $\bv_{\bk}$ and the charge $e$ have opposite signs for the
particle and the hole components.  The real-space motion, however, is
sensitive to the particle-hole mixing since the particle and the hole
move in opposite directions. This leads to a modified real-space equation of
motion
\begin{equation}\label{semi2}
\dot{\br}=(|u_{\bk}|^2-|v_{\bk}|^2)\bv_\bk,
\end{equation}
 which represents the net center of mass motion of the bogolon wavepacket.

The semiclassical approximation amounts to introducing periodic time
dependence in the Hamiltonian, in lieu of the magnetic
field: ${\cal H}(t+T)={\cal H}(t)\equiv{\cal H}_{\bk(t)}$.  Unlike in the gapless
case where ${\cal H}(t)=$const, here ${\cal H}(t)$ exhibits time
dependence due to
the momentum-dependent gap $\Delta(t)=\tilde\Delta_{\bk(t)}$, while
$\epsilon=\epsilon_\bk$ remains a constant of motion.  To find the
solution of the time-dependent Schr\"{o}dinger equation,
\begin{equation}\label{schr}
 i\hbar\dot\Psi={\cal H}(t)\Psi,
\end{equation}
we employ the Floquet theorem\cite{stockmann}, which is the analog of the
familiar Bloch theorem for time-periodic Hamiltonians; it states that
solutions of Eq.\ (\ref{schr}) have the form
$\Psi(t)=e^{-i\tilde{E}t/\hbar}f_E(t)$ with $f_E(t+T)=f_E(t)$. Here
$e^{-i\tilde{E}T/\hbar}$ and $f_E(T)$ are the eigenvalue and the eigenstate,
respectively, of the Floquet operator
\begin{equation}\label{Floquet}
{\cal F} = {\cal T} \exp\left[{-i\over \hbar} \int_0^T {\cal H}(t)dt\right],
\end{equation}
and ${\cal T}$ represents the time-ordering operator.  If we regard
the two-component structure of $\Psi(t)$ as a pseudospin, then
the Floquet states precess about a time dependent axis $\epsilon
{\bf \hat z} + \Delta(t){\bf\hat x}$.

The quantity $\tilde{E}$ has dimensions of energy and is closely related to
the quasiparticle energy of the original time-independent
problem. Since the Floquet equation yields $e^{-i\tilde{E}T/\hbar}$ and
$T=2\pi/\omega_c$ it is clear that $\tilde{E}$ is defined only modulo
$\hbar\omega_c$. This is analogous to momentum being defined only
modulo reciprocal lattice vectors in the Bloch theory. Henceforth we refer
to $\tilde{E}$ as `quasi-energy'.

In order to obtain analytic results we simplify our model further. We
assume a free-electron dispersion
$\epsilon_\bk=\hbar^2k^2/2m-\epsilon_F$ and that the FAM gap is piecewise
constant and dependent only on the momentum direction. We take
$\tilde\Delta_\bk$ equal to zero on the arcs and $\pm\Delta_0$ elsewhere as
illustrated in Fig.\ \ref{fig1}d. The details of the calculation of the Floquet
quasi-energy are provided in the Methods. Here we present the results
and discuss their implications.

The quasi-energy as a function of both the band energy $\epsilon$ and
the magnetic field is shown in Fig.\ \ref{quasienergy}a.  It contains
most of the physics of this model.  Fig.\ \ref{quasienergy}b shows a
cut along the $\epsilon$ direction for constant $B$.  The
quasiparticle energy dispersion is obtained by a simple `unwinding'
procedure, described in Methods. Energy bands
separated by small gaps result, in close analogy to the Bloch energy
bands. The density of states in Fig.\ \ref{quasienergy}c displays
clear periodic structure with frequency that can be estimated from
Eq.\ (\ref{floqE}) as
\begin{equation}\label{fe}
F_\epsilon = {\theta_a \over \pi\hbar\omega_c}.
\end{equation}
This is in agreement with the exact numerical results which are
discussed in the next section.

We now turn our attention to the low-energy behavior of the FAM as a function of
field $B$.  Near the Fermi energy, $\epsilon\rightarrow 0$, the quasi-energy coincides with the
quasi-particle energy, $E=\tilde{E}$, and no unwinding is necessary. When
$\epsilon \to 0$ the quasi-energy vanishes, but the density of states, which is
proportional to $dE/d\epsilon$, depends strongly on the magnetic field.
In Fig.\ \ref{quasienergy}d we present the quasi-energy for a few
different values of $\epsilon$ close to the Fermi energy as a function
of $1/B$.  In certain magnetic fields the density of lines is high and
this translates to the sharp peaks in the DOS shown in Fig.\ \ref{quasienergy}e . This result is directly
related to the experimentally observed oscillations.  From Eq.\
(\ref{floqE}) we may deduce that a peak in the Fermi energy DOS occurs whenever
$\cos(\Delta_0\theta_g/\hbar\omega_c)=0$, leading to oscillation
frequency
\begin{equation}\label{fb}
F = \Delta_0{\theta_g m_c \over \pi\hbar e}.
\end{equation}
Using $\Delta_0=80$meV, $\theta_g = \pi/4$ (Ref.\ \onlinecite{damascelli1})
and a cyclotron mass $m_c =3m_e$ (Ref.\ \onlinecite{bangura1}) we estimate $F=518$T, very close to the dominant experimental
frequency 530-540T in YBCO.

\begin{figure*}
\includegraphics[width=\columnwidth]{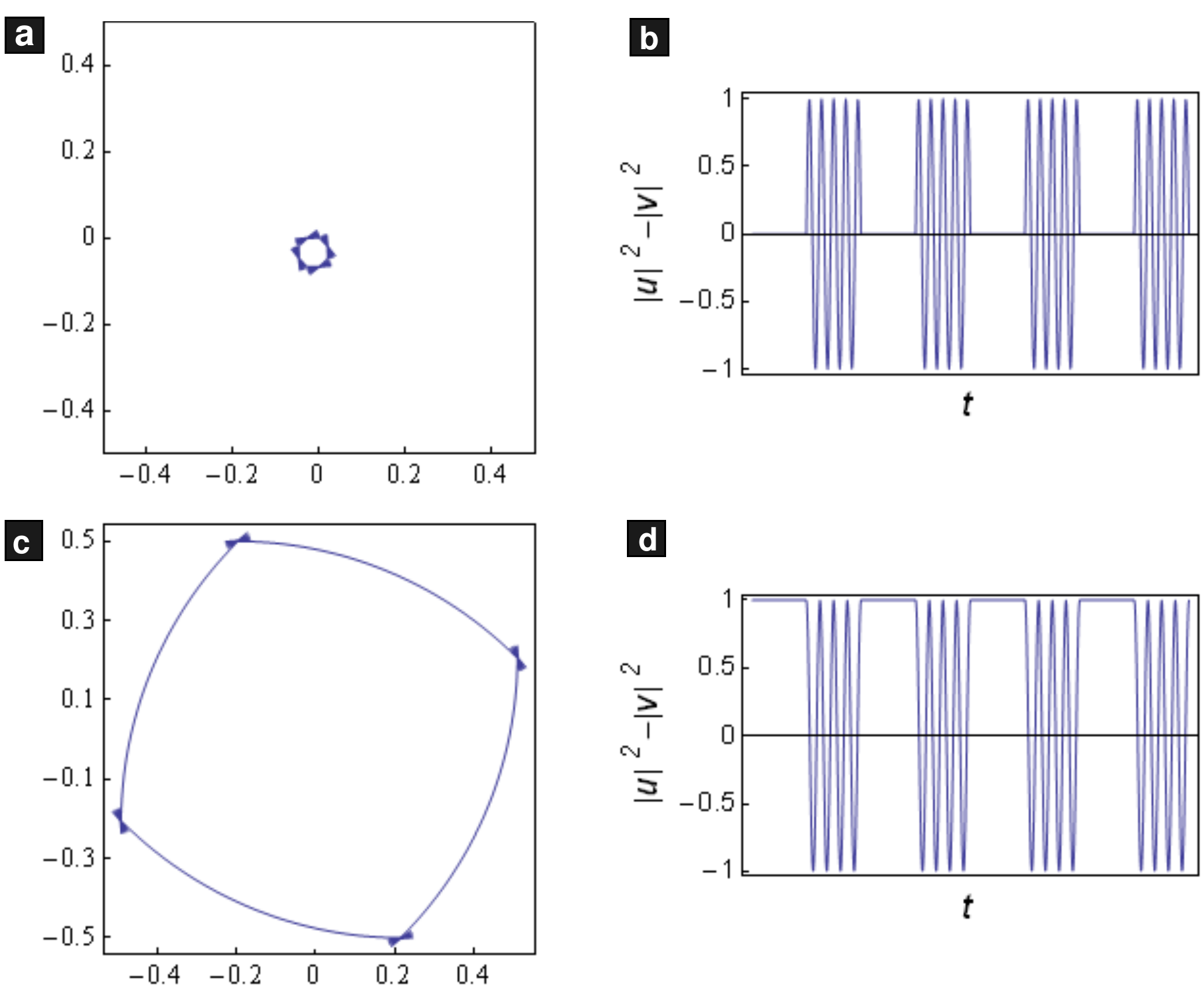}
\caption{{\bf Semiclassical trajectories.} Left column: Sample real-space
trajectories. Right column: the quasi-particle group velocity relative to $v_F$, $|u|^2-|v|^2$, as a function of time along the trajectory.  When $\epsilon \ll \Delta_0$ the bogolon pseudospin lies in the $y-z$ plane.  Its
projection on the $z$ axis is $|u|^2-|v|^2$.  The states contributing
to high DOS [panels (a) and (b) with $\hbar\omega_c = {\Delta_0\theta_g
/(4+{1\over2})\pi}$] have their pseudospin pointing in the $y$-direction on the
arc and are precessing in the $y-z$ plane in the gapped region.  This
means that on the arc the state is a
perfect mixture of a particle and a hole, resembling a familiar Andreev
bound state. The associated real-space
trajectory encloses a particularly small area, since on the arc, the group velocity vanishes, $|u|^2-|v|^2=0$. On the other hand, the states near the DOS minimum [panels (c) and (d), $\hbar\omega_c =
{\Delta_0\theta_g /4\pi}$], point in the $z$ pseudo-spin direction and are
thus {\it either} particles {\it or} holes.  Their motion is dominated by
the arc region, and the real-space trajectory encloses a large area.
The DOS oscillations with $B$ may be understood as follows. In order to
exhibit no real-space motion on the arc, Andreev states must point in
the $y$ pseudospin direction.  In the gapped region, the pseudospin
precesses about the $x$-axis (assuming $\epsilon \ll \Delta_0$). If
the magnetic field is such that the time to traverse the gapped
region, $\theta_g/\omega_c$, allows exactly an integer plus one half cycles,
the pseudospin will end up in the $-y$ direction and will continue to
have zero motion on the next arc, thus giving a consistent Floquet state.
}\label{trajectories}
\end{figure*}

Some intuitive understanding of the origin of the DOS oscillations can be
gained by examining  a typical low-energy Floquet state $f_E(t)$  and its
associated real-space trajectory. This is illustrated in Fig.\
\ref{trajectories}. The states contributing to high DOS are reminiscent
of the Andreev
bound states found on extended impurities and on sample edges in
$d$-wave superconductors\cite{adagideli}.
The periodicity in inverse magnetic field in this model is a
consequence of the periodic appearance of these Andreev-type states on the
Fermi arcs at low energies.

The quantum oscillation mechanism described above is dominant for
quasiparticle energies much smaller than the gap amplitude $\Delta_0$,
and does not involve the conventional Onsager-Lifshitz action
considerations. Obviously, the action must be considered to obtain the
normal-metal behavior in the limit of small gap or large arcs. The
interplay of the two mechanisms then becomes complicated and we
briefly discuss it in Methods.

\section{Lattice model}

\begin{figure*}
\includegraphics[width = 14.0cm]{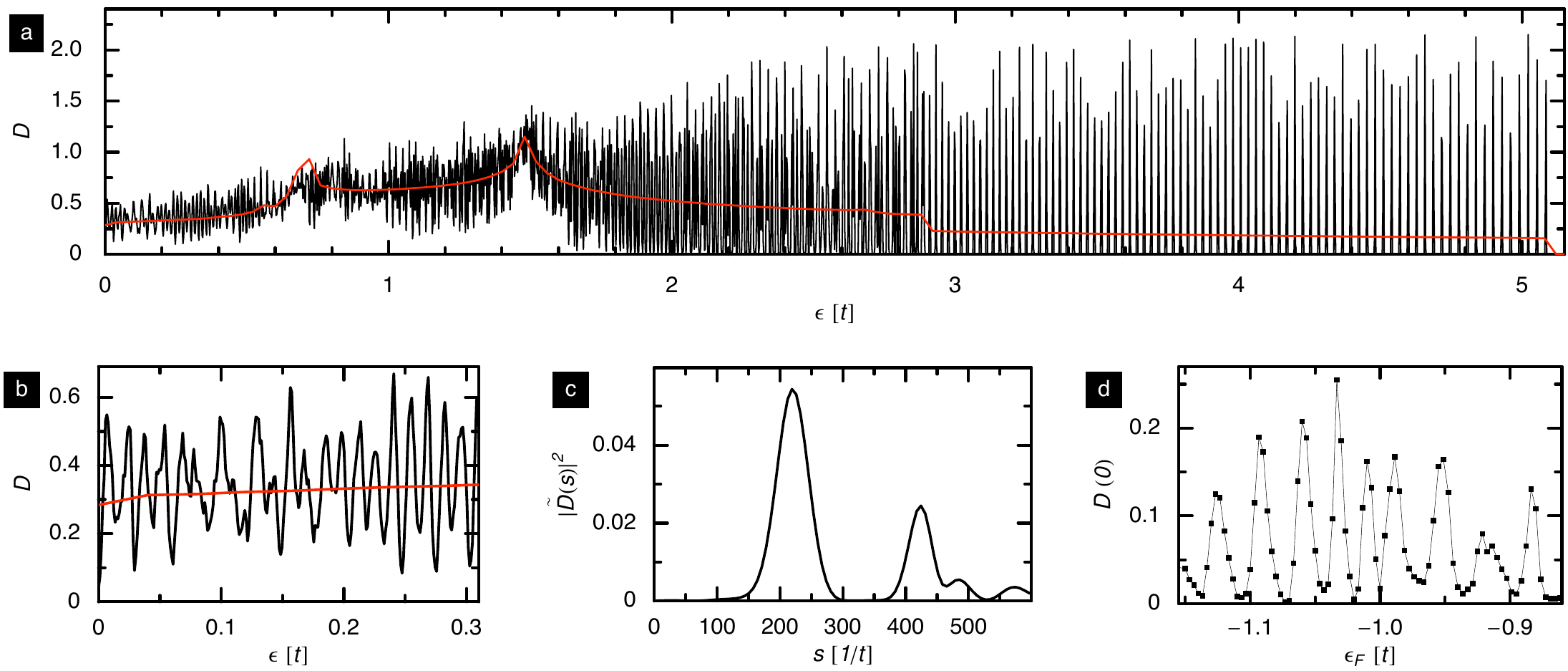}
\caption{{\bf Exact diagonalization of the lattice model. }
a) DOS as a function of energy in the Fermi-arc metal in zero (red)
and non-zero (black) magnetic field corresponding to two vortices in a
$20\times 20$ magnetic unit cell. In YBCO with lattice constant
$a_0\simeq 4$\AA \ this corresponds to the physical field of about
64T. The parameters used are as follows, $\Delta_0/t=1$,
$\epsilon_F/t=-1.3$, $\nu=0.6$ and $\tau=0.1$. b) The
low-energy DOS for the same parameters, in detail. c) The power spectrum of the
low-energy DOS showing dominant frequency of oscillations $210t^{-1}$
and its second harmonic. d) DOS at the Fermi level as a function of
$\epsilon_F$.
}
\label{fig_FT}
\end{figure*}
In order to exemplify the validity of our semiclassical analysis we now consider
a fully quantum-mechanical lattice model of the Fermi-arc metal and confirm the
existence of quantum oscillations by exact numerical calculations.
To this end, we study the real-space version of the Hamiltonian (\ref{h1}),
\begin{equation}\label{h2}
{\cal H}=\sum_{ij\sigma}t_{ij}e^{i\theta_{ij}} c^\dagger_{i\sigma}c^{}_{j\sigma}
+\sum_{ij}\left[\tilde\Delta^{}(\br_i,\br_j)\ c^\dagger_{i\uparrow}c^{\dagger}_{j\downarrow}+{\rm h.c.}\right],
\end{equation}
where $c^\dagger_{i\sigma}$ creates an electron with spin $\sigma$ at site $\br_i$
of the square lattice. The effect of magnetic field is described by the
usual Peierls factors
$\theta_{ij}=(2\pi/\Phi_0)\int_{\br_i}^{\br_j}\bA\cdot{\bf dl}$
and the SC order parameter $\tilde\Delta^{}(\br_i,\br_j)$ is correspondingly
taken to contain a periodic lattice of Abrikosov vortices. This is achieved by
adopting
\begin{equation}\label{dr}
\tilde\Delta^ {}(\br_1,\br_2)=e^{i\Theta(\bR)}\sum_\bk e^{i\br\cdot\bk}\tilde\Delta_\bk,
\end{equation}
where $\bR=(\br_1+\br_2)/2$ is the center of mass and $\br=\br_1-\br_2$ the
relative coordinate of the Cooper pair. The vortex lattice is reflected in the
phase $\Theta(\bR)$ winding by $2\pi$ around each vortex while the Fermi-arc
structure in $k$-space follows from taking $\tilde\Delta_\bk$ as described by
Eq.\ (\ref{dk}). In order to keep subsequent calculations simple we limit
ourselves to the nearest-neighbor electron hopping in the kinetic term:
$t_{ii}=\epsilon_F$, $t_{ij}=-t$ for $(i,j)$ nearest neighbors and $t_{ij}=0$
otherwise.

We solve the problem posed by Eqs.\ (\ref{h2},\ref{dr}) by employing the
Franz-Tesanovic (FT) transformation\cite{ft1}. This unitary transformation
removes the non-trivial phase $\Theta(\bR)$ from the pairing term and renders
the transformed Hamiltonian translationally invariant with a unit cell
containing two superconducting vortices. Eigenstates of this new Hamiltonian
then can be conveniently found by appealing to the Bloch theorem.  Ref.\
\onlinecite{vft1} gives a detailed description of the implementation of the
FT transformation to the lattice model of $s$-, $p$- and $d$-wave
superconductors. The treatment of our modified $d$-wave SC follows as a
straightforward generalization of this procedure.

The results of our numerical calculations are summarized in Fig.\
\ref{fig_FT}. The density of states $D(\epsilon)$ of the modified
$d$-wave SC in the applied magnetic field shows the expected Landau
level structure\cite{vft1} at energies exceeding about
$3\Delta_0$. The surprising new result is the appearance of a clear
periodic structure in $D(\epsilon)$ even at {\em low energies} inside
the SC gap. This is unlike the ordinary $d$-wave SC where the Landau
level type oscillations are known to be absent at low energies
\cite{ft1,vft1,kita1,marinelli1} (we have confirmed this
result by setting $\nu=0$ in our model). We attribute the low-energy
oscillations to the gapless regions of the BZ implied by the modified
$d$-wave order parameter Eq.\ (\ref{dk}) with $\nu\neq 0$.

The power spectrum of the low-energy DOS displayed in Fig.\
\ref{fig_FT}c confirms the periodic structure with a period that scales
with $1/B$, in accordance with the semiclassical result Eq.\ (\ref{fe}). Physical observables, such as the specific heat and
resistivity, depend on DOS at the Fermi level, $D(\epsilon_F)$. Fig.\
\ref{fig_FT}d shows that $D(\epsilon_F)$ exhibits similar oscillatory
behavior.

The above exact numerical results show unambiguous evidence for quantum
oscillations in FAM, a system that by construction exhibits genuine
Fermi arcs terminated by a pairing gap. Although the exact
diagonalization technique does not allow us to study these
oscillations as a function of smoothly varying field $B$ (and thus
compare directly to experiment), we have verified that oscillations in
the energy variable are in qualitative as well as semi-quantitative
agreement with the semiclassical picture presented in
Sec. II. Specifically, we have analyzed the dependence of oscillation
frequency $s$ on the gap amplitude $\Delta_0$ and the arc length $\nu$
and found these in agreement with the semiclassical predictions. This
comparison is discussed more fully in the Methods.

\section{Outlook}

The original observation of quantum oscillations in YBCO \cite{taillefer1} has
been interpreted as quantitatively consistent with the
ARPES measurements \cite{shen1} by assuming that the Fermi
arc observed in ARPES was a part of a Fermi pocket resulting from the Fermi
surface reconstruction due to a symmetry breaking instability with wavevector
$(\pi,\pi)$. Such a Fermi pocket would have an area of about 2\% of
the BZ, consistent with the observed oscillation frequency. The ARPES data used
for this comparison however pertain not to YBCO but to a different
high-$T_c$ compound Na$_{2-x}$Ca$_x$Cu$_2$O$_2$Cl$_2$ (NaCCOC).
If one uses instead the ARPES data on YBCO that became available more recently
\cite{damascelli1}, then the agreement disappears:
the corresponding Fermi pocket would comprise about $5\%$ of the BZ, leading to
the frequency more than twice that observed in experiment. The two
experiments are easily reconciled by appealing to the Fermi arc picture advocated
above, where the frequency of oscillations does not relate to any Fermi surface area but originates from the periodic appearance of Andreev-type states associated with Fermi arcs.  The oscillation frequency (\ref{fb}) depends on the gap amplitude and the size of the gapped region in the Fermi-arc metal.

The above comparison illustrates some key differences between the conventional
Onsager-Lifshitz picture of quantum oscillations in YBCO
\cite{millis1,kee1,rice1} and the mechanism invoking genuine
Fermi arcs terminated by a pairing gap proposed in this study. The latter
relies only on the FS structure that is directly seen in ARPES while the former
must make assumptions about unseen portions of the FS in the parts of the BZ
where the ARPES indicates a large gap\cite{millis1,kee1,rice1}. A direct
observation of
Fermi arcs at low temperatures in high magnetic fields would discriminate
between the two pictures. Under such conditions ARPES experiments are not
feasible but it should be possible to image the FS by means of the scanning
tunneling probe using the technique of Fourier-transform interference
spectroscopy \cite{hoffman1,mcelroy1,tami1}. Even if confirmed, the
microscopic origin of the Fermi arc phenomenon remains an open question, the
answer to which may pave the road towards the full solution of the cuprate
mystery.

\section{Acknowledgments}
The authors acknowledge illuminating discussions with D.\ Bonn, W.\ Hardy,
B.\ Seradjeh, L.\ Taillefer, Z.\ Tesanovic, O.\ Vafek, M.\ Vojta and N-C. Yeh.
The work was supported in part by NSERC, CIfAR (MF), DFG through SFB 608 (HW), the
Packard Foundation, and the Research Corporation (GR).

\section{Methods}

\subsection{Experimental considerations}
A crucial assumption we made in deducing the
oscillatory behavior of $D(\epsilon)$ in Eqs.\ (\ref{fe},\ref{fb})
is that the gap structure itself is field-independent. This is
equivalent to assuming that magnetic field enters the underlying mean-field
Hamiltonian via the usual minimum substitution but does not alter any of its
parameters. This known to be true in
ordinary metals but it is less obvious that this assumption applies to our
Fermi-arc metal which relies on the existence of the residual SC gap.
The amplitude and the
$k$-space structure of the latter could be susceptible to magnetic field.
Specifically, if the Fermi arc length depended on $B$ then the frequency of the quantum oscillations would itself
become a function of $B$, in contradiction to experimental finding of constant
frequency. Since there exist no independent measures of the Fermi arc length as
a function of $B$, and since the microscopic theory underlying
the Fermi arc phenomenon is also unknown, we must adopt the arc length
independence on $B$ in the experimentally relevant interval 20-60T as an
additional assumption of our phenomenological model.

\subsection{Floquet eigenstates}

 With a piecewise constant gap $\Delta(t)$ and a constant band energy
$\epsilon$ the time ordered exponent of the Floquet operator
(\ref{Floquet}) may be written as a product of 8 operators, ${\cal F}
= U_-U_0U_+U_0U_-U_0U_+U_0$.  These correspond to the time evolution
operators on the eight segments of the Fermi surface (4 arcs and 4
gapped segments),
\begin{equation}\label{Floquet2}
U_0 = e^{-i\epsilon\sigma_3T_0/\hbar}, \;\;\;\; U_{\pm} = e^{-i(\epsilon\sigma_3\pm\Delta_0\sigma_1)T_1/\hbar}
\end{equation}
where $T_0=\theta_a/\omega_c$ and $T_1=\theta_g/\omega_c$ are the times to traverse the arc and the gapped regions, respectively.   The properties of the Pauli matrices allow us
to write
\begin{eqnarray}
U_0 &=& \cos\left({\epsilon\theta_a\over \hbar\omega_c}\right)-i\sin\left({\epsilon\theta_a\over \hbar\omega_c}\right)\sigma_3  \\
U_{\pm}&=& \cos\left({E_{\bk}\theta_g\over \hbar\omega_c}\right)-
i\sin\left({E_{\bk}\theta_g\over \hbar\omega_c}\right){\epsilon \sigma_3\pm\Delta_0\sigma_1 \over E_{\bk}},
\nonumber
\end{eqnarray}
with $E_{\bk} = \sqrt{\epsilon^2+\Delta_0^2}$.  Since the motion on
the first 4 segments is the same as on the last 4 segments we may define
${\cal F}_{1/2}^2 \equiv {\cal F}$ and find the eigenstates of half
the cycle.  Multiplying the four operators yields
\begin{equation}
{\cal F}_{1/2} = \begin{pmatrix} A & iB \\ iB^* & A^* \end{pmatrix}
\end{equation}
with
\begin{eqnarray}
A &=& \left( {\Delta_0 \over E_{\bk}} \sin{\alpha_1}\right)^2 + e^{-2i\alpha_0}\left(\cos{\alpha_1}-i{\epsilon \over E_{\bk}}\sin{\alpha_1}\right)^2 \nonumber \\
B &=&  {\Delta_0 \over E_{\bk}}e^{-i\alpha_0}
\sin{\alpha_1}\left({\epsilon\over
 E_{\bk}}\sin{\alpha_1} \cos{\alpha_0} + \cos{\alpha_1} \sin{\alpha_0}\right), \nonumber
\end{eqnarray}
where $\alpha_0 = \epsilon\theta_a/ \hbar\omega_c$ and $\alpha_1 =
E_{\bk}\theta_g / \hbar\omega_c$.  Using the fact that ${\cal
F}_{1/2}$ is unitary (and therefore $|A|^2+|B|^2=1$) we find the real part of its
eigenvalue as cosine of the phase $\phi = (2\pi \tilde{E}/\hbar\omega_c)$ and deduce the quasi-energy
\begin{equation}\label{floqE}
\tilde{E} = {\hbar\omega_c \over \pi}\arccos{[{\rm Re}(A)]}
\end{equation}
This  is presented in Fig.\ \ref{quasienergy}a.

\subsection{Energy unwinding}
The full quasiparticle energy, $E$, is deduced from the quasienergy as a function of the band energy, $\tilde{E}(\epsilon)$, through the following unwinding procedure.  First, since the inverse cosine function in the quasienergy gives angles between $0$ and $\pi$ we interpret decreasing segments of the quasienergy as resulting from angles between $\pi$ and $2\pi$.  On these segments we replace: $\arccos[{\rm Re}(A)]\to 2\pi -\arccos[{\rm Re}(A)]$.  The result are disconnected monotonically increasing segments of energy between $0$ and $\hbar\omega_c$, the energy bands.  In analogy with Bloch bands, each band begins and ends with a flat dispersion.
We are free to shift the energy by an integer multiple of $\hbar\omega_c$ and do so uniformly in each band, i.e., without 'breaking' the bands.  Different bands are shifted by different amounts in order to create a monotonically increasing function.  We define $E=\tilde{E}+n\hbar\omega_c$ where $n$ is the band unwinding index.
In the high energy region, $\epsilon \gg \Delta_0$, the SC gap is negligible and we expect the energy to converge to the band energy.  Indeed, if we choose the band unwinding index to simply count the bands starting at zero for the first band, we obtain a monotonic function with small gaps separating the bands at low $\epsilon$.  At high $\epsilon$ we recover the desired linear dependence with slope 1.  However, a small offset between $E(\epsilon)$ and  $\epsilon$ appears.  We interpret this offset as requiring larger gaps at low energy.  Usually steps of $2\hbar\omega_c$ or $3\hbar\omega_c$ are sufficient. We note that with the exception of vanishing arc length, the slope at $\epsilon=0$ is finite, meaning that the Fermi energy is a band midpoint where a gap does not open.  In addition, we expect the energy to be antisymmetric with respect to $\epsilon$ and therefore  $\tilde{E}(\epsilon=0)=0$.  As a result, the lowest energy band is not shifted.

\begin{figure*}
\includegraphics[width = 7.0cm]{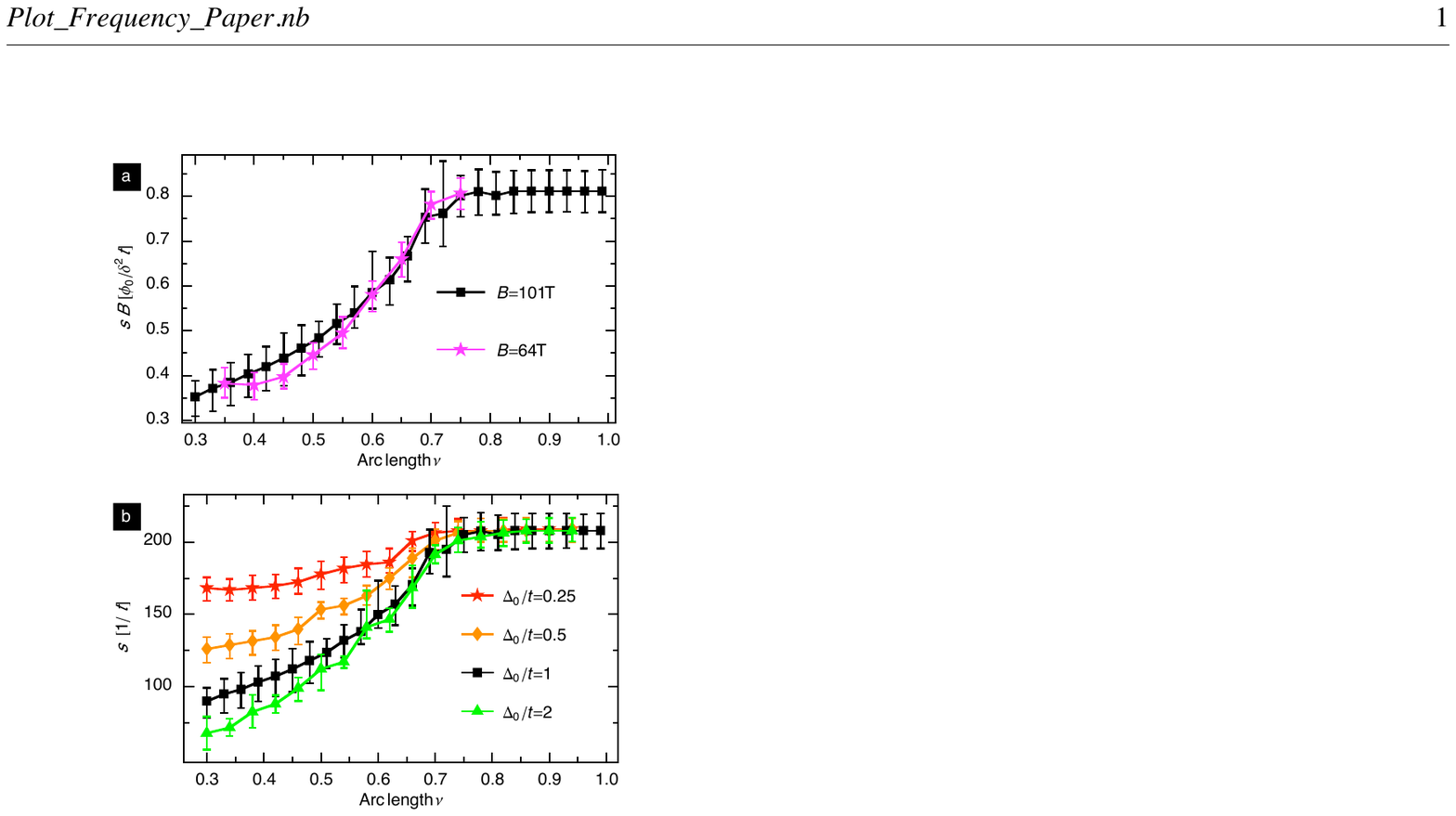}
\caption{{\bf Oscillation frequency in the lattice model.} The frequency
$s$ is extracted from the power spectrum analysis of oscillations in
the low-energy quasiparticle DOS as a function of the arc length
$\nu$ with a fixed chemical potential. Panel (a) shows scaling with magnetic field, obtained by
considering $16\times 16$ and $20\times 20$ magnetic unit cell. Panel
(b) shows the dependence on the gap amplitude at constant field $B$.
}
\label{fig5}
\end{figure*}
%

\subsection{Action considerations}

The quantum oscillations described in Sec. II do not originate
from and are not affected by the action quantization.  Nevertheless,
the action plays an important role in the limit of small gap or large
arcs, when Andreev bound states are rare.  We use the circular
momentum-space trajectory and the Floquet state pseudospin to
determine the real-space paths and to calculate the associated action.  At
energies close to $\epsilon_F$ the action is proportional to the
energy.  To a good approximation $S/\hbar \propto l_B^2 {\cal
A}_k|\cos(\Delta_0 \theta_g/\hbar\omega_c)|$ with the magnetic length $l_B$ and ${\cal
A}_k$ the area in $k$-space bounded by the trajectory.  Thus, the linear
dependence of the action on $1/B$ is modulated by $|\cos(\Delta_0
\theta_g/\hbar\omega_c)|$, the projection of the pseudospin on the
$z$-direction while on the arc.  Therefore, for magnetic fields with
peaked DOS the action is zero.  This means that the Andreev
states are allowed by the action quantization.  The action slope close
to these points is very large (about 3.5 larger than the slope
associated with the full Fermi surface area) so that many other states at near magnetic fields
are allowed by the action quantization.  The fact that more states are
allowed does not change the DOS periodicity in $1/B$, however it may
broaden the observed oscillation frequency range.  When the gapped
region shrinks (arc length increases) or the gap amplitude decreases
the frequency of DOS modulations due to the Andreev states will
decrease.  The action quantization then becomes more important.
A second frequency which reflects the full Fermi surface area
will appear and eventually dominate when the gap closes (the cosine is
simply 1 in this limit).  In this way our model recovers the usual
Onsager-Lifshitz quantization in the limit of vanishing gap.

\subsection{Frequency analysis}

We have analyzed the behavior of the oscillation frequency $s$ in the
lattice model of Sec. III as a function of magnetic field $B$, arc
length $\nu$ and maximum gap size $\Delta_0$ (at a fixed chemical potential).
Our results are
summarized in Fig.\ \ref{fig5}. The oscillation frequency scales
linearly with $1/B$ as expected on the basis of semiclassical
Eq. (\ref{fe}). For small and intermediate arc length $\nu$ it
increases linearly with $\nu$, also in accord with
Eq. (\ref{fe}). Panel (b) shows that the frequency also exhibits
dependence of the gap amplitude, not expected on the basis of
Eq. (\ref{fe}). This dependence, however, is seen to saturate in the
limit of large $\Delta_0$. This is consistent with the fact that
Eq. (\ref{fe}) is valid only in the regime $\epsilon\ll\Delta_0$ and
this condition is not fully met in the lattice model for small
$\Delta_0/t$ since we must consider a finite energy window to extract
the oscillation frequency. A more careful analysis of Eq.\
\ref{floqE} indeed yields a weak gap dependence of the oscillation
frequency $F_\epsilon$ with a trend resembling that displayed in Fig.\
\ref{fig5}b.


\end{document}